\documentclass[a4paper]{jpconf}
\usepackage{amsmath,amssymb}
\usepackage{graphicx}  
\renewcommand{\thefootnote}{\fnsymbol{footnote}}

\newcommand{\p}{\partial}
\newcommand{\pslash}{p\kern-1ex /}
\newcommand{\lslash}{l\kern-1ex /}
\newcommand{\kslash}{k\kern-1ex /}
\newcommand{\dslash}{\p\kern-1.2ex /}
\newcommand{\Dslash}{{\cal D}\kern-1.5ex /}
\newcommand{\Aslash}{A\kern-1.2ex /}

\newcommand{\re}{{\rm Re}}

\newcommand{\Dodwf}{\mathcal{D}}
\newcommand{\bea}{\begin{eqnarray}}
\newcommand{\eea}{\end{eqnarray}}

\newcommand{\EQ}{\hspace{-2mm} &=& \hspace{-2mm}}

\newcommand{\BAN}{\begin{eqnarray*}}
\newcommand{\EAN}{\end{eqnarray*}}
\begin{document}
\title{Simulation of lattice QCD with domain-wall fermions}

\author{Ting-Wai Chiu (for the TWQCD Collaboration)}

\address{Physics Department, National Taiwan University, Taipei 106, Taiwan}
\address{Center for Quantum Science and Engineering, National Taiwan University, Taipei 106, Taiwan}

\ead{twchiu@phys.ntu.edu.tw}

\begin{abstract}
Quantum Chromodynamics (QCD) is the fundamental theory  
for the interaction between quarks and gluons.  
It manifests as the short-range strong interaction 
inside the nucleus and plays an important role in the 
evolution of the early universe, from the quark-gluon 
phase to the hadron phase. 
To solve QCD is a grand challenge, 
since it requires very large-scale numerical simulations 
of the discretized action of QCD on the 4-dimensional 
space-time lattice.
Moreover, since quarks are relativistic fermions, 
the fifth dimension is introduced such that massless quarks 
with exact chiral symmetry can be realized at finite lattice 
spacing, on the boundaries of the fifth dimension, the so-called 
domain-wall fermion (DWF).  
In this work, I discuss how to simulate lattice QCD with DWF 
such that the chiral symmetry can be preserved optimally with 
a finite extent in the fifth dimension. I also outline the simulations
which have been performed by the TWQCD Collaboration and present 
some recent physical results.    

\end{abstract}



\section{Introduction}

Quantum Chromodynamics (QCD) is the fundamental theory  
for the interaction between quarks and gluons.
It provides the theoretical framework to understand 
the nuclear force/energy from the first principles. 
Moreover, QCD plays an important role in the 
evolution of the early universe, from the quark-gluon 
phase to the hadron phase. Since quarks are relativistic fermions, 
they possess chiral symmetry in the massless limit.   
Chiral symmetry forbids additive mass renormalization 
which causes the fine-tuning problem associated with the scalar field.
In QCD, the chiral symmetry [$ SU_L(N_f) \times SU_R(N_f) $]
of $ N_f $ massless quarks is spontaneously broken to
$ SU_V(N_f) $, due to the strong interaction between
quarks and gluons. This gives the (nearly) massless Goldstone
bosons (pions) and their specific interactions.
To investigate the spontaneous chiral symmetry breaking
as well as hadron physics from the first principles of QCD, 
it requires nonperturbative methods.
So far, lattice QCD is the most promising approach, 
discretizing the continuum space-time on a 4-dimensional lattice \cite{Wilson:1974sk}, 
and computing physical observables by Monte Carlo simulations \cite{Creutz:1980zw}.
However, in lattice QCD, formulating lattice fermions
with exact chiral symmetry at finite lattice spacing is rather nontrivial. This 
is realized through domain-wall fermions (DWF) on the (4+1)-dimensional
lattice \cite{Kaplan:1992bt,Kaplan:sg}
and overlap fermions on the 4-dimensional lattice
\cite{Neuberger:1997fp,Narayanan:1995gw}.

Lattice QCD with exact chiral symmetry \cite{Kaplan:1992bt,Neuberger:1997fp}
is an ideal theoretical framework to study
nonperturbative physics from the first principles of QCD.
However, it is rather nontrivial to perform Monte Carlo simulations 
such that the chiral symmetry is preserved at high precision 
and all topological sectors are sampled ergodically. 

Currently, there are three groups (RBC/UKQCD, JLQCD, TWQCD) around the world 
performing large-scale dynamical simulations of lattice QCD with exact chiral symmetry. 
Since the computational requirement for these dynamical simulations is 
$10-100$ times that of their counterparts using traditional lattice fermions 
(e.g., Wilson fermions, staggered fermions, and their variants),   
they are often performed with state-of-the-art architectures. 
While RBC/UKQCD and JLQCD have been using IBM Blue Gene supercomputers, 
TWQCD has been using a GPU cluster since 2009 (currently consisting 
of $ 320 $ Nvidia GPUs, with sustained $ 100 $ Tflops/s). 

The RBC/UKQCD Collaborations have been using conventional 
domain-wall fermions with the Shamir kernel \cite{Shamir:1993zy, Allton:2007hx}, 
which suffers from large chiral symmetry breaking (i.e., large residual mass), 
especially in the finite temperature QCD.
On the other hand, the JLQCD Collaboration used overlap fermions 
in a fixed topology \cite{Fukaya:2006vs}, 
which attains very good chiral symmetry but 
at the expense of sampling all topological sectors.  
To overcome the deficiencies of the above two approaches, 
the TWQCD Collaboration has been using 
optimal domain-wall fermions (ODWF) \cite{Chiu:2002ir-Chiu:2003bv,Chiu:2009wh}
to preserve the chiral symmetry, which not only attains a good chiral symmetry
with a modest extension (e.g., $N_s=16$) in the fifth dimension, 
but also samples all topological sectors ergodically.
Recently, the JLQCD Collaboration has started a new project \cite{JLQCD} to use 
conventional domain-wall fermions with the scaled Shamir kernel, 
performing dynamical simulations with more than 6 racks of the IBM Blue Gene/Q supercomputer.
In other words, now all 3 groups (RBC/UKQCD, JLQCD, TWQCD)
are using domain-wall fermions (DWF) to perform large-scale dynamical 
simulations of lattice QCD.   

In this work, we discuss how to simulate lattice QCD with DWF 
such that the chiral symmetry can be preserved optimally with 
a finite extent in the fifth dimension. We also outline the simulations
which have been performing by the TWQCD Collaboration and present 
some recent physical results.    

Mathematically, optimal domain-wall fermions (ODWF) is a theoretical framework 
to preserve the chiral symmetry maximally with a set of analytical weights, 
$ \{ \omega_s, s = 1, \cdots, N_s \} $, 
one for each layer in the fifth dimension \cite{Chiu:2002ir}. 
Thus the artifacts due to the chiral
symmetry breaking with finite $ N_s $ can be reduced to the minimum, 
especially in the chiral regime.
In general, the 4-dimensional effective Dirac operator of massless ODWF 
can be written as \cite{Chen:2012jya}
\bea
\label{eq:odwf_4d}
\begin{aligned}
D &= \frac{1}{2r} [1+ \gamma_5 S_{opt}(H) ], 
\hspace{5mm}
S_{opt}(H) = \frac{1-\prod_{s=1}^{N_s} T_s}{1 + \prod_{s=1}^{N_s} T_s}, \\
T_s &= \frac{1-\omega_s H}{1+\omega_s H},   
\hspace{5mm}
H = c H_w ( 1 + d \gamma_5 H_w)^{-1}, 
\hspace{5mm}
r = [2 m_0 (1-d m_0) ]^{-1},  
\end{aligned}
\eea
where $ c $ and $ d $ are constants, and 
$ H_w = \gamma_5 D_w $, with $ D_w $ the usual Wilson-Dirac operator
plus a negative parameter $ -m_0 \; ( 0 < m_0 < 2 ) $. 
Here $ S_{opt}(H) = H R_Z(H) $, where $ R_Z(H)$ is the Zolotarev optimal 
rational approximation of $ (H^2)^{-1/2} $ \cite{Chiu:2002eh}. 

Recently we have demonstrated that it is feasible to perform a large-scale dynamical QCD 
simulation with ODWF, which not only preserves the chiral symmetry to a good precision, 
but also samples all topological sectors ergodically \cite{Chiu:2011dz}.   
To recapitulate, we perform HMC simulations of 2 flavors QCD on a $ 16^3 \times 32 $ lattice, 
with ODWF at $ N_s = 16 $ and plaquette gauge action at $ \beta = 5.95 $. 
Then we compute the low-lying eigenmodes of the overlap Dirac operator 
and use its index to obtain the topological charge of each 
gauge configuration, from which we compute the topological susceptibility
for 8 sea-quark masses. Our result of the topological susceptibility agrees with the 
sea-quark mass dependence predicted by the NLO ChPT \cite{Mao:2009sy}, 
and provides the first determination of both the pion decay constant 
($ F_\pi = \mathrm{92(12)(2)~ MeV} $)
and the chiral condensate 
($\Sigma^{\overline{\rm MS}}\mathrm{(2~GeV)=[259(6)(7)~MeV]^3} $) 
simultaneously from the topological susceptibility. 
Furthermore, our recent results of the mass and the decay constant 
of the pseudoscalar meson \cite{Chiu:2011bm}
are also in good agreement with 
the sea-quark mass dependence predicted by NLO ChPT \cite{Gasser:1984gg}, 
from which we obtain
the low-energy constants $ F $, $ \Sigma $, $ \bar{l}_3 $ and $ \bar{l}_4 $.
With the low-energy constants, we determine the average up and down quark mass 
($m_{ud}^{\overline{\rm MS}}(\mathrm{2~GeV}) = \mathrm{4.17(13)(19)~MeV}$), 
and the chiral condensate ($\Sigma^{\overline{\rm MS}}\mathrm{(2~GeV)=[230(4)(6)~MeV]^3} $). 
Our results of the topological susceptibility together with the mass and decay constant
of the pseudoscalar meson assert that the nonperturbative chiral dynamics 
of the sea-quarks are well under control in our HMC simulations. 

Recently we have extended our simulations to two sets of larger lattices: 
(i) $ 20^3 \times 40 \times 16 $ with plaquette gauge action at $ \beta = 5.95 $, 
for 6 sea-quark masses corresponding to pion masses in the range 230-450 MeV; 
(ii) $ 24^3 \times 48 \times 16 $ with plaquette gauge action at $ \beta = 5.95 $, 
for 4 sea-quark masses corresponding to pion masses in the range 230-450 MeV.
Now, after simulating 2-flavors QCD on various lattices 
($ 16^3 \times 32 $, $ 20^3 \times 40 $, $ 24^3 \times 48 $), 
we are ready to perform dynamical simulations of $(2+1)$-flavors and $(2+1+1)$-flavors 
QCD on the $ 32^3 \times 64 \times 16 $ lattice, with pion mass close to the physical value.  
Our strategy will be outlined in the final section.

\section{Theoretical aspects of domain-wall fermions}

In general, for a given $ N_s $ (the number of sites in the fifth dimension), 
the (mathematically) maximal chiral symmetry can be attained by 
the optimal domain-wall fermion (ODWF) \cite{Chiu:2002ir} with the operator
\bea
\label{eq:D_odwf_def}
[\Dodwf(m_q)]_{xx';ss'} = 
  (\rho_s D_w + 1)_{xx'} \delta_{ss'}
 +(\sigma_s D_w - 1)_{xx'} L_{ss'},       
\eea
where $ \rho_s = c \omega_s + d $, $ \sigma_s = c \omega_s - d $, and $ c $, $d$ are constants.
The indices $ x $ and $ x' $ denote the sites on the 4-dimensional space-time lattice, 
while $ s $ and $ s' $ the layers in the fifth dimension. Here $D_w$ is the standard Wilson Dirac operator 
plus a negative parameter $-m_0 \; (0 < m_0 < 2)$,
\begin{equation}
(D_w)_{xx'} = -\frac{1}{2} \sum_{\mu} \left[
  (1-\gamma_\mu)U_\mu(x)\delta_{x+\hat{\mu},x'}
 +(1+\gamma_\mu)U^\dagger_\mu(x')\delta_{x-\hat{\mu},x'} \right]
 + (4 - m_0),
\end{equation}
where $U_\mu(x)$ denotes the link variable pointing from $ x $ to $ x + \hat\mu $,  
\begin{equation}
L = P_+ L_+ + P_- L_-, \quad P_\pm = (1\pm \gamma_5)/2,
\end{equation}
and
\begin{equation}
(L_+)_{ss'} = \left\{ 
    \begin{array}{ll} 
      - r m_q \delta_{N_s,s'}, & s = 1,  \quad r \equiv 1/[2m_0(1-dm_0)] \\ 
      \delta_{s-1,s'}, & 1 < s \leq N_s  
    \end{array}\right.,
\quad L_-=(L_+)^{T}.
\end{equation}
The weights $ \{ \omega_s \} $ along the fifth dimension are 
fixed according to the formula derived in \cite{Chiu:2002ir} such 
that the maximal chiral symmetry is attained. 
In general, for other DWFs without maximal chiral symmetry, 
the weights $\{\rho_s\}$ and $\{\sigma_s\}$ have different values. 
For example, for the conventional (Shamir) DWF, which 
has been used by the RBC/UKQCD Collaborations, $c=d=1/2 $ and $ \omega_s = 1, \forall s $;
for the scaled Shamir DWF (with scaling factor = 2), which is being used 
by the JLQCD Collaboration, $ c = 1 $, $ d = 1/2 $, and $ \omega_s = 1, \forall s $.   

The breaking of chiral symmetry due to finite $N_s$ in the fifth dimension can be measured
by the residual mass emerging in the axial Ward identity on the lattice. 
For the general DWF operator (\ref{eq:D_odwf_def}), the axial Ward identity is derived
in~\cite{Chen:2012jya}, and a new formula for the residual mass is also obtained, 
\bea
m_{res} = 
\left< \frac{ \tr(D_c + m_q)^{-1}_{0,0} }{ \tr[(D_c^\dagger + m_q)(D_c+m_q)]^{-1}_{0,0} } \right>_{\{U\}} - m_q,
\label{eq:mres}
\eea
where $ (D_c + m_q)^{-1} $ denotes the valence quark propagator with $ m_q $ equal to the sea-quark mass,
tr denotes the trace running over the color and Dirac indices, and the subscript $ \{U\} $ denotes averaging
over an ensemble of gauge configurations. 
Formula (\ref{eq:mres}) is useful in practice, since it immediately gives the 
residual mass once the 12 columns of the quark propagator are computed.
Moreover, an upper-bound for the residual mass of ODWF is derived in~\cite{Chen:2012jya}.
It asserts that for $ N_s $ less than some threshold value ($\sim 16-18 $), the residual mass is 
an exponentially decay function of $ N_s $. This implies that ODWF can provide a viable way 
to preserve chiral symmetry to a good precision (e.g., $ m_{res} a \sim 10^{-5} $) 
with a modest $ N_s $ (e.g., $ N_s \simeq 16 $).

\section{HMC simulation of lattice QCD with domain-wall quarks}

First of all, we perform the even-odd preconditioning on the DWF operator (\ref{eq:D_odwf_def}), 
which is essential for lowering the condition number as well as halving the memory consumption. 
Since $D_w$ commutes with $(\rho)_{ss'} \equiv \rho_s \delta_{ss'}$ and 
$(\sigma)_{ss'} = \sigma_s \delta_{ss'} $, (\ref{eq:D_odwf_def}) becomes 
\begin{equation}
\Dodwf(m_q)=D_w(\rho+\sigma L)+(1-L) = D_w [ c \omega (1+L) + d (1-L)] + (1-L), 
\label{eq:D_odwf_def2}
\end{equation}
where $ (\omega)_{s,s'} = \omega_s \delta_{s,s'}$ is a diagonal matrix in the fifth dimension.  
Now, separating the even and the odd sites on the 4D space-time lattice, 
(\ref{eq:D_odwf_def2}) can be written as
\bea
\Dodwf(m_q) =
\begin{pmatrix}
4 - m_0 & D_w^{\text{EO}} \\
D_w^{\text{OE}} & 4 - m_0 
\end{pmatrix} 
[c\omega(1+L)+ d(1-L)] + (1-L)  
=
\begin{pmatrix}
X & D_w^{\text{EO}} Y \\
D_w^{\text{OE}} Y & X
\end{pmatrix},
\label{eq:D_odwf_eo}
\eea
where
\begin{equation}
X \equiv (4 - m_0 )[ c \omega(1+ L)+d(1-L)] + (1-L), \quad Y \equiv c \omega (1+L) + d(1-L).  
\end{equation}
Now we further rewrite it in a more symmetric form by defining
\begin{equation}
\label{eq:m5}
M_5\equiv \omega^{-1/2} YX^{-1} \omega^{1/2}
= \left\{(4-m_0) + \omega^{-1/2} \left[ c (1+L)(1-L)^{-1}  + d \omega^{-1} \right]^{-1} \omega^{-1/2} \right\}^{-1},
\end{equation}
and
\begin{equation}
S_1\equiv \omega^{-1/2} YX^{-1} = M_5 \omega^{-1/2},
\quad S_2\equiv Y^{-1} \omega^{1/2}.
\end{equation}
Then (\ref{eq:D_odwf_eo}) becomes
\begin{equation}
\Dodwf(m_q) =  
S_1^{-1}
\begin{pmatrix}
1 & M_5 D_w^{\text{EO}}  \\
M_5 D_w^{\text{OE}} & 1 
\end{pmatrix}
S_2^{-1}
=S_1^{-1}
\begin{pmatrix}
1 & 0 \\
M_5 D_w^{\text{OE}} & 1
\end{pmatrix}
\begin{pmatrix}
1 & 0 \\
0 & C
\end{pmatrix}
\begin{pmatrix}
1 & M_5 D_w^{\text{EO}} \\
0 & 1
\end{pmatrix}
S_2^{-1},
\label{eq:D_odwf_decomp}
\end{equation}
where the Schur decomposition has been used in the last equality, with the Schur complement 
\begin{equation}
\label{eq:C_def}
C \equiv 1 - M_5 D_w^{\text{OE}} M_5 D_w^{\text{EO}}.
\end{equation}
Note that the quark mass dependence $(m \equiv r m_q)$ only resides in $M_5$ through $L$.

Since $ \det\Dodwf = \det S_1^{-1} \cdot \det C \cdot \det S_2^{-1} $, and
$ S_1 $ and $ S_2 $ do not depend on the gauge field, we can just use $ C $
in the Monte Carlo simulation. After including the Pauli-Villars fields (with $ m_q = 1/r $), 
the pseudofermion action for 2 flavors QCD (in the isospin limit $ m_u = m_d $) can be written as
\bea
\label{eq:Spf}
S_{pf} = \phi^\dagger C_1^\dagger ( C C^\dagger)^{-1} C_1 \phi, \quad C_1 \equiv C(m_q = 1/r), 
\eea
where $ \phi $ and $ \phi^\dagger $ are complex scalar fields carrying the same quantum numbers 
(color, spin) of the quark fields. Including the gluon fields, the partition function for 2 flavors QCD 
can be written as
\bea
\label{eq:Z_nf2}
Z = \int[dU][d\phi^{\dag}][d\phi]\exp\left(-S_g[U]-\phi^\dagger C_1^\dagger ( C C^\dagger)^{-1} C_1 \phi \right), 
\eea
where $S_g[U]$ is the lattice action for the gauge field, e.g., the Wilson plaquette action
\bea
S_g[U]=\beta\sum_{plaq.}\left\{1-\frac{1}{3} \re \Tr (U_p) \right\}, \hspace{4mm} \beta=\frac{6}{g^{2}}.
\eea
It is rather difficult to simulate (\ref{eq:Z_nf2}) directly with the Metropolis algorithm. 
The conventional wisdom to handle this problem is to 
introduce a fictitious dynamics to guide the Monte Carlo simulation, 
i.e., the Hybrid Monte Carlo (HMC) \cite{Duane:1987de}.
Since the pseudofermion action (\ref{eq:Spf}) is positive-definite, $ \phi $ can be generated by 
the heat-bath method with Gaussian noise $\eta$ satisfying the Gaussian distribution 
$\exp(-\eta^\dagger\eta)$. That is, to solve the following equation with 
the conjugate gradient algorithm    
\bea
\label{eq:CG noise}
C_1 \phi= C \eta \Leftrightarrow C_1^\dagger C_1 \phi= C_1^\dagger C \eta.
\eea
Then the fictitious molecular dynamics only involves the gauge fields 
$ \left\{ A_l \right\} $ and their conjugate momenta $ \left\{ P_l \right\} $,
where $A_l = A_l^a t^a$ is the matrix-valued gauge field corresponding
to the link variable $U_l = \exp(iA_l^a t^a)$.
The Hamiltonian of the molecular dynamics is
\bea
\mathcal{H}=\frac{1}{2}\sum_{l,a}(P_l^a)^2+S_g[U]+ \phi^\dagger C_1^\dagger ( C C^\dagger)^{-1} C_1 \phi, 
\eea
and the partition function can be written as
\bea
{\cal Z}=\int[dU][dP][d\phi][d\phi^{\dag}]\exp(-\mathcal{H}).
\eea
The Hamilton equations for the fictitious molecular dynamics are
\bea
\label{eq:molecular_U}
\frac{dA_l^a(\tau)}{d\tau} \EQ \frac{\partial\mathcal{H}}{\partial P_l^a(\tau)}=P_l^a(\tau)
\Leftrightarrow \frac{dU_l(\tau)}{d\tau}=iP_l(\tau)U_l(\tau), \\
\label{eq:molecular_P}
\frac{dP_l^a(\tau)}{d\tau} \EQ -\frac{\partial\mathcal{H}}{\partial A_l^a(\tau)}
=-\frac{\partial S_g}{\partial A_l^a(\tau)}-\frac{\partial S_{pf}}{\partial A_l^a(\tau)}. 
\eea
These two equations together imply that $ d\mathcal{H}/d\tau = 0 $, which gives
\bea
\label{eq:molecular_P_a}
P_l^a\frac{dP_l^a(\tau)}{d\tau}=-\frac{dS_g}{d\tau}-\frac{dS_{pf}}{d\tau},
\eea
as an alternative form of (\ref{eq:molecular_P}). 

The algorithm of HMC simulation can be outlined as follows:
\begin{enumerate}
\item Choose an initial gauge configuration $\{U_l\}$.
\item Generate $P_l^a$ with Gaussian weight $\exp(\{P_l^{a}\}^2/2)$.
\item Generate $\eta$ with Gaussian weight $\exp(-\eta^\dagger\eta)$.
\item Compute $\phi $ according to (\ref{eq:CG noise}).
\item With $\{\phi\}$ held fixed, integrate (\ref{eq:molecular_U}) and (\ref{eq:molecular_P})
      by an algorithm (leapfrog/Omelyan) which ensures exact reversibility and area-preserving map 
      in the phase space for any $\delta\tau$.
\item Accept the new configuration $\{U_l' \}$ generated by the molecular dynamics with probability
      $min(1,e^{-\Delta\mathcal{H}})$, where $\Delta\mathcal{H}\equiv\mathcal{H}(U_l^{\prime},P_l^{\prime})-\mathcal{H}(U,P)$.
      This completes one HMC trajectory. 
\item For the next trajectory, go to (ii).  
\end{enumerate}

To summarize, we first generate the random noise vector $ \eta $ with Gaussian distribution,
$ \exp(-\eta^\dagger \eta) $, then we obtain $ \phi = C_1^{-1} C \eta $ using the conjugate gradient (CG).
With fixed $ \phi $, the system is evolved under a fictitious Hamiltonian dynamics,
the so-called molecular dynamics (MD). In the MD, we use the Omelyan integrator 
\cite{Omelyan:2001,Omelyan:2002,Omelyan:2003},
and the Sexton-Weingarten multiple-time scale method \cite{Sexton:1992nu}.
The most time-consuming part in the MD is to compute the fermion force  
$ -\partial S_{pf}/\partial A_l^a(\tau) $ which is required for updating the conjugate momenta 
in~(\ref{eq:molecular_P}), since it involves solving the linear system 
$ (C C^\dagger) v =  C_1 \phi $ with CG. Here we take advantage of 
the remarkable floating-point capability of the Nvidia GPU and perform the 
CG with mixed precision \cite{Chiu:2011rc}.
Moreover, the computations of the gauge force and the update of the gauge field 
are also performed in GPUs. In other words, the entire HMC trajectory is computed by GPUs.  
Furthermore, we introduce an auxiliary heavy fermion field with mass $ m_H $ ($ m_q \ll m_H < 1/r $)
similar to the case of Wilson fermions \cite{Hasenbusch:2001ne}.
For two-flavors QCD, the pseudofermion action (with $ C_H \equiv C(m_H) $) becomes,  
\BAN
S_{pf}^H = \phi^{\dagger} C_H^{\dagger} (CC^{\dagger})^{-1} C_H \phi + 
           \phi_H^{\dagger} C_{PV}^{\dagger} (C_H C_H^{\dagger})^{-1} C_{PV} \phi_H, 
\EAN
which gives exactly the same fermion determinant of (\ref{eq:Spf}).   
Nevertheless, the presence of the heavy fermion field plays a crucial role in 
reducing the light fermion force and its fluctuation, thus diminishes the change 
of the Hamiltonian in the MD trajectory, and enhances the acceptance rate.

In two-flavors QCD, only the quantum fluctuations (internal fermion loops) of the 
lightest $ u $ and $ d $ quarks (in the isospin limit $ m_u = m_d $) are incorporated,  
while those of the heavier quarks ($ s $, $ c $, $ b $ and $ t $) are neglected.   
To incorporate the dynamical $ s $ quark, one must use a pseudofermion action 
different from that of two-flavors QCD with degenerate masses (\ref{eq:Spf}). 
A straightforward approach is to take the inverse square root of the quark matrix 
(and the quarter-root of the Pauli-Villars matrix) of the two-flavors QCD, namely
\BAN
S_{pf}^{N_f=1} = \phi^\dagger (C_1^\dagger C_1)^{1/4} (C C^\dagger)^{-1/2} (C_1^\dagger C_1)^{1/4} \phi.
\EAN
However, the inverse square root and the quarter-root operations cannot be evaluated exactly  
since the cost of solving the eigenproblem of $ C C^\dagger $ or $ C_1^\dagger C_1 $ is prohibitive.
A way to resolve this problem is to use the optimal rational approximations for 
$ (C C^\dagger)^{-1/2} $ and $ (C_1^\dagger C_1)^{1/4} $ respectively,  
similarly to the rational hybrid Monte Carlo (RHMC) \cite{Clark:2006fx}. 

Recently, Chen and Chiu have constructed a novel DWF action for one flavor QCD \cite{Chen:2013of}, which is exact,  
without taking the square root, and amenable to HMC simulation. 
This novel pseudofermion action for one flavor QCD can be written as 
\bea
\label{eq:Spf_Nf1}
S_{pf}^{N_f=1}
&=& \left(0  \hspace{4mm}  \phi_1^{\dag} \right)
        \left[I-c_- v_{-}^{T}\omega^{-1/2} \frac{1}{H(m)} \omega^{-1/2}v_{-}\right]
        \left(\begin{array}{c}
           0  \\
        \phi_1
        \end{array}\right)+ \nonumber\\
&  & \left(\phi_2^{\dag}  \hspace{4mm} 0\right)
        \left[I+c_+ v_{+}^{T}\omega^{-1/2} \frac{1}{H(1) - \Delta_+(m)P_+} \omega^{-1/2}v_{+}\right]
        \left(\begin{array}{c}
        \phi_2 \\
           0
        \end{array}\right), 
\eea 
where $ \phi_1 $ and $ \phi_2 $ are complex scalar fields on the 4-dimensional space-time lattice,
with color and 2-spinor indices, and 
\BAN
\label{eq:H}
H(m) &=& \gamma_5 R_5 \left[ D_w + \omega^{-1/2}\left\{\omega^{-1}d+c[1+L(m)][1-L(m)]^{-1} \right\}^{-1} \omega^{-1/2} \right] \\
&\equiv& \gamma_5 R_5 \left[ D_w + M_{+} P_{+} + M_{-} P_{-} \right], \hspace{4mm} (R_5)_{s,s'} = \delta_{s',N_s+1-s}, \\
\Delta_{\pm}(m) &=& R_5[M_{\pm}(1)-M_{\pm}(m)]. 
\EAN
The coefficients $ c_{\pm} $ and the vectors $ v_{\pm} $ in (\ref{eq:Spf_Nf1}) 
only depend on the parameters $ c $, $ d $, and $ \{ \omega_s, s=1,\cdots, N_s \} $, 
and they will be shown explicitly in Ref. \cite{Chen:2013of}.  
A detailed description of our HMC simulations will be presented in a forthcoming paper.

\section{Gauge ensembles}

For zero temperature QCD, we have been working on the following ensembles of two-flavors QCD with ODWF:  
\begin{itemize}
\item $ 16^3 \times 32 $ ($ a \sim 0.1 $~fm), with plaquette gauge action at $ \beta = 5.95 $, 
for 8 sea-quark masses corresponding to pion masses in the range 230-560 MeV \cite{Chiu:2011dz, Chiu:2011bm}.   
\item $ 20^3 \times 40 $ ($ a \sim 0.1 $~fm), with plaquette gauge action at $ \beta = 5.95 $, 
for 6 sea-quark masses corresponding to pion masses in the range 230-450 MeV \cite{Chiu:2012jm}.
\item $ 24^3 \times 48 $ ($ a \sim 0.1 $~fm) with plaquette gauge action at $ \beta = 5.95 $, 
for 4 sea-quark masses corresponding to pion masses in the range 230-450 MeV.
\end{itemize}
The first two lattice sizes ($ 16^3 \times 32 $, and $ 20^3 \times 40 $) have been completed, 
while the third lattice size ($ 24^3 \times 48 $) is expected to be completed in the summer of 2013.
   
For finite temperature QCD, we have been working on the following ensembles of two-flavors QCD with ODWF:  
\begin{itemize}
\item $ 16^3 \times 6 $, with plaquette gauge action, for 30-40 $ \beta $ values in the range $ [5.00, 5.95] $,   
each of 3-4 sea-quark masses.   
\item $ 24^3 \times 8 $, with plaquette gauge action, for 30-40 $ \beta $ values in the range $ [5.00, 5.95] $, 
each of 3-4 sea-quark masses.   
\end{itemize}
The first lattice size ($ 16^3 \times 6 $) has been completed, 
while the lattice size ($ 24^3 \times 8 $) is still in progress.  

For the quark part, we use ODWF with $ c = 1 $, $ d = 0 $ (i.e., $ H = H_w $), 
$ N_s = 16 $, and $ \lambda_{min}/\lambda_{max} = (0.01,0.02,0.05)/6.2 $, 
where different values of $ \lambda_{min} $ have been used for different gauge ensembles.  
In general, for each sea-quark mass, we generate the initial 400-500 trajectories on a single GPU.
After discarding 200-300 trajectories for thermalization, we sample one configuration
every 5 trajectories, resulting in 20-32 ``seed" configurations for each sea-quark mass. 
Then we use these seed configurations as the initial configurations 
for independent simulations on 20-32 GPUs.     
Each GPU generates 200-250 trajectories independently.  
Then we accumulate a total of $ 5000 $ trajectories for each sea-quark mass. 
From the saturation of the binning error of the plaquette, as well as
the evolution of the topological charge, 
we estimate the autocorrelation time to be around 10 trajectories. 
Thus we sample one configuration every 10 trajectories, 
and obtain $ 500 $ configurations for each sea-quark mass.

\begin{figure*}[tb]
\begin{center}
\begin{tabular}{@{}c@{}c@{}}
\includegraphics*[width=8cm,clip=true]{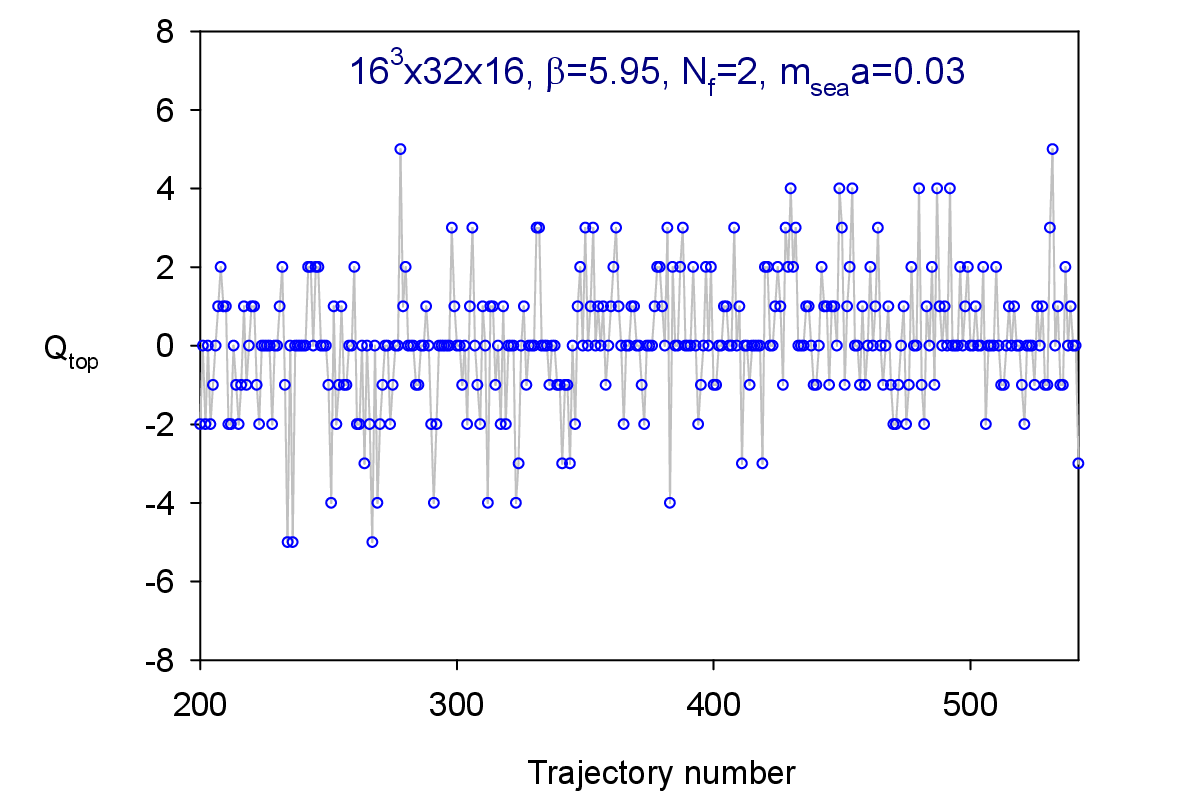}
&
\includegraphics*[width=8cm,clip=true]{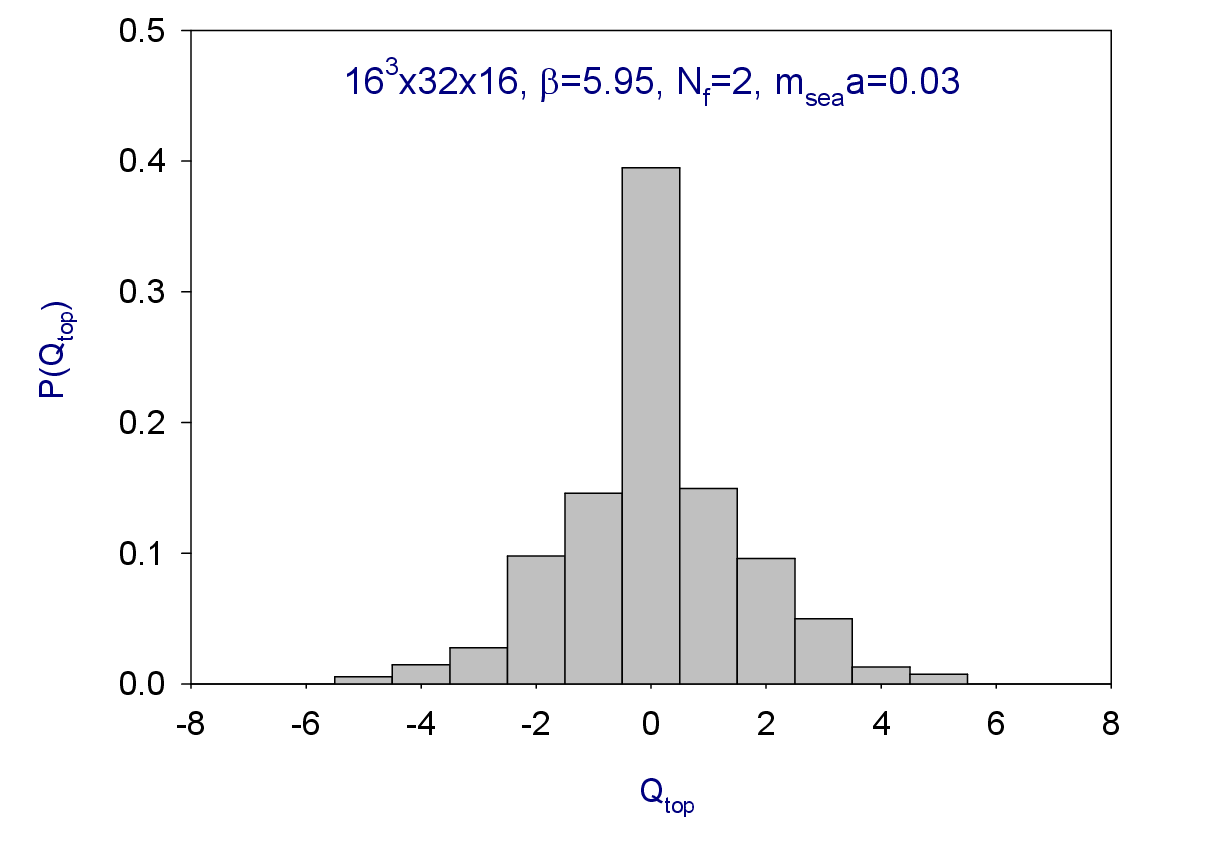}
\\ (a) & (b)
\end{tabular}
\caption{(a) The topological charge versus the trajectory in the HMC simulation of two-flavors QCD with ODWF.
             The lattice is $ 16^3 \times 32 \times 16 $ with the spatial box $ \sim (\mbox{2 fm})^3 $ 
             and the sea-quark mass corresponding to $ M_\pi \sim 360 $ MeV.
             The line connecting the data points is only for guiding the eyes.
             (b) The histogram of the topological charge distribution in (a).}
\label{fig:Qt_history}
\end{center}
\end{figure*}

\begin{figure*}[tb]
\begin{center}
\begin{tabular}{@{}c@{}c@{}}
\includegraphics*[width=8cm,clip=true]{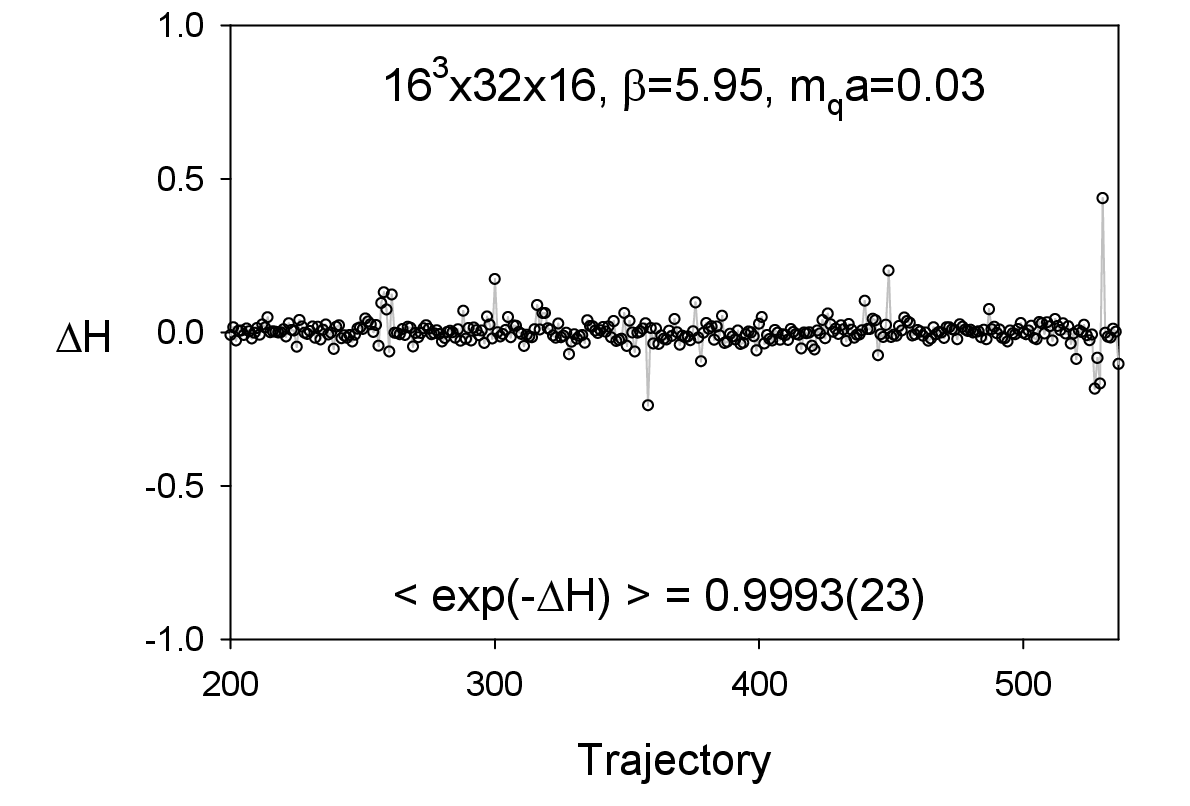}
&
\includegraphics*[width=8cm,clip=true]{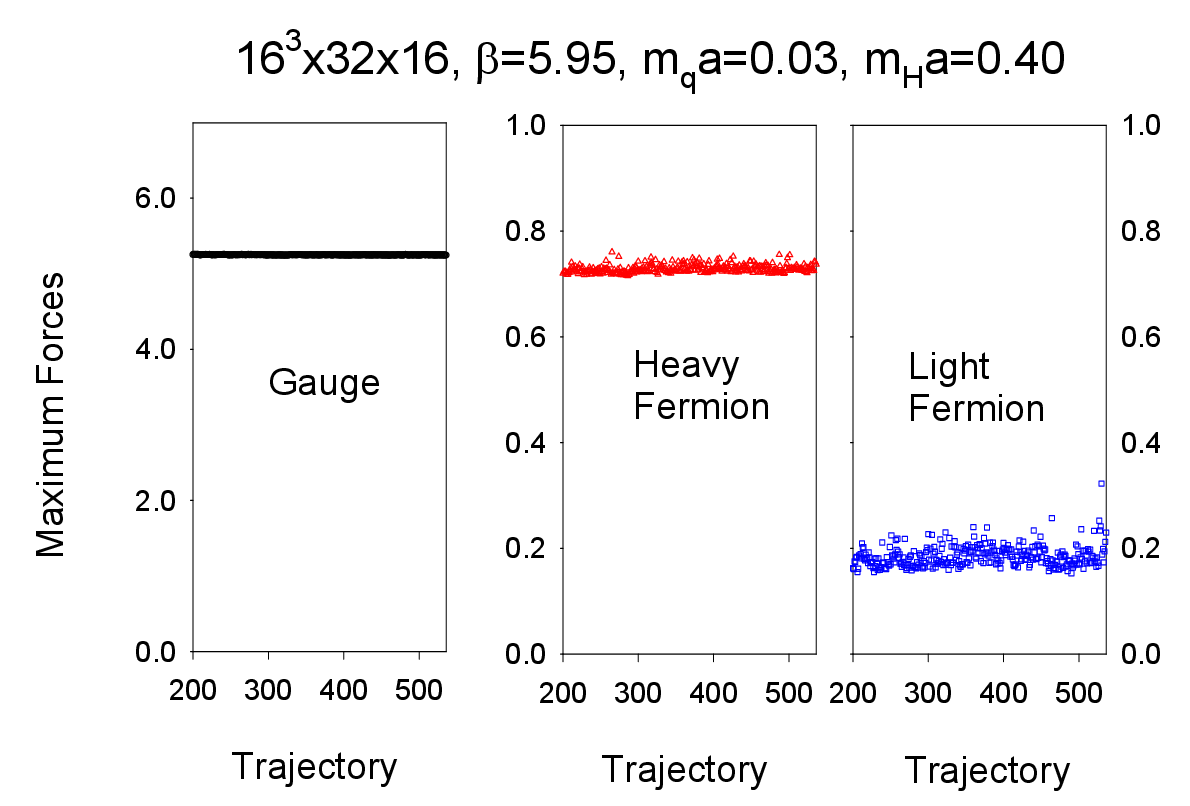}
\\ (a) & (b)
\end{tabular}
\caption{(a) The $ \Delta {\cal H} $ versus the trajectory in the HMC simulation of two-flavors QCD with ODWF.
             The lattice is $ 16^3 \times 32 \times 16 $ with the spatial box $ \sim (\mbox{2 fm})^3 $, 
             and the sea-quark mass corresponding to $ M_\pi \sim 360 $ MeV.
             The line connecting the data points is only for guiding the eyes.
             (b) The maximum forces of the gauge field, heavy fermion field, and light fermion field.}
\label{fig:HMC_features}
\end{center}
\end{figure*}

For each configuration, we calculate the zero modes plus 80-180
conjugate pairs of the lowest-lying eigenmodes of the overlap
Dirac operator. We outline our procedures as follows. First, we
project 250 low-lying eigenmodes of $ H_w^2 $, 
using adaptive thick restart Lanczos algorithm (a-TRLan), where each eigenmode
has a residual less than $ 10^{-12} $. Then we approximate the sign
function of the overlap operator by the Zolotarev optimal rational
approximation with 64 poles, where the coefficients are fixed
with $ \lambda^2_{max} = (6.4)^2 $ and $ \lambda_{min}^2 $
equal to the maximum of the 250 projected eigenvalues of $ H_w^2 $.
Then the sign function error is less than $ 10^{-14} $. 
Using the 250 low-modes of $ H_w^2 $ and the Zolotarev approximation
with 64 poles, we use the a-TRLan algorithm again to
project the zero modes plus 80-180 conjugate pairs of the lowest-lying
eigenmodes of the overlap operator, where each eigenmode has 
a residual less than $ 10^{-12} $. We store all projected eigenmodes
for the later use. 

In figures~\ref{fig:Qt_history}-\ref{fig:HMC_features},   
we present the essential characteristics of our HMC simulations,  
which are quite universal for any lattice sizes and sea-quark masses. 
In figure~\ref{fig:Qt_history} (a), we plot the topological charge versus 
the trajectory in the HMC simulation of two-flavors QCD with ODWF.
The lattice is $ 16^3 \times 32 \times 16 $ with the spatial box $ \sim (\mbox{2 fm})^3 $ 
and the sea-quark mass corresponding to $ M_\pi \sim 360 $ MeV. 
All these HMC trajectories are computed in one GPU, with a total of 543 trajectories, 
and the initial 200 trajectories are discarded for thermalization.   
We see that the topological charge has a short auto-correlation time, 
and each topological sector is sampled ergodically. 
Moreover, the topological charge distribution behaves like a Gaussian distribution
with the mean close to zero, as shown in figure~\ref{fig:Qt_history} (b).
In figure~\ref{fig:HMC_features} (a), we plot the change of the Hamiltonian 
of each trajectory, which is quite smooth, without any spikes in all trajectories. 
Using the measured value of $ \left< \Delta {\cal H} \right> = 0.00167(240) $, 
we estimate the theoretical acceptance rate 
$ P_{\rm HMC} = {\rm erfc} \left( \sqrt{ \left< \Delta {\cal H} \right>}/2 \right) = 0.977(10)$,
which agrees with the measured acceptance rate $ 0.983(7) $.
Moreover, the measured value of $ \left< \exp(-\Delta {\cal H}) \right> = 0.9993(23) $, 
in good agreement with the condition $ \left< \exp(-\Delta {\cal H}) \right> = 1 $
which follows from the area-preserving property of the HMC simulation.    
In figure~\ref{fig:HMC_features} (b), we plot the maximum force (averaged over all links) 
among all momentum updates in a trajectory, for the gauge field, the heavy fermion field, 
and the light fermion field respectively. The forces all behave smoothly for all trajectories.

\section{Some recent results}

To verify whether a dynamical simulation of lattice QCD captures
the nonperturbative chiral dynamics of the sea-quarks, a prerequisite is to  
examine to what extent the pion mass ($ M_\pi $) and decay constant ($ F_\pi $)
agree with the sea-quark mass dependence as predicted by the next-to-leading-order
chiral perturbation theory (NLO ChPT) \cite{Gasser:1984gg}
and to check whether the resulting low-energy constants are reasonable or not. 
Furthermore, to check whether a dynamical simulation of lattice QCD samples
all topological sectors ergodically, it is necessary to measure the topological 
susceptibility versus the sea-quark mass and to compare the results  
with the sea-quark mass dependence as predicted by the NLO ChPT \cite{Mao:2009sy}.
We have performed these tests for the gauge ensembles on the 
$ 16^3 \times 32 \times 16 $ lattice \cite{Chiu:2011dz,Chiu:2011bm}. 

In this section, we present preliminary results of these tests for the 
gauge ensembles on the $ 20^3 \times 40 \times 16 $ lattice \cite{Chiu:2012jm}.

Using the valence quark propagator with quark mass equal to the sea-quark mass, 
we compute the time-correlation function of the pseudoscalar interpolator
\BAN
\label{eq:Gt}
C_\pi(t) = \sum_{ \vec{x} } \tr \{ \gamma_5 ( D_c + m_q )^{-1}_{0,x} \gamma_5
                ( D_c + m_q )^{-1}_{x,0} \},
\EAN
where the trace runs over the Dirac and color space.
Then the ensemble average $ \langle C_\pi(t) \rangle $ of each $ m_q $ 
is fitted to the formula
$ Z [ e^{-M_{\pi} a t} + e^{-M_{\pi} a (T-t)} ] / (2 M_{\pi} a) $
to extract the pion mass $ M_{\pi} a $ and the decay constant
$ F_{\pi} a = m_q a \sqrt{2Z}/(M_{\pi} a)^2 $. 

\begin{figure*}[tb]
\begin{center}
\begin{tabular}{@{}c@{}c@{}}
\includegraphics*[width=7.5cm,clip=true]{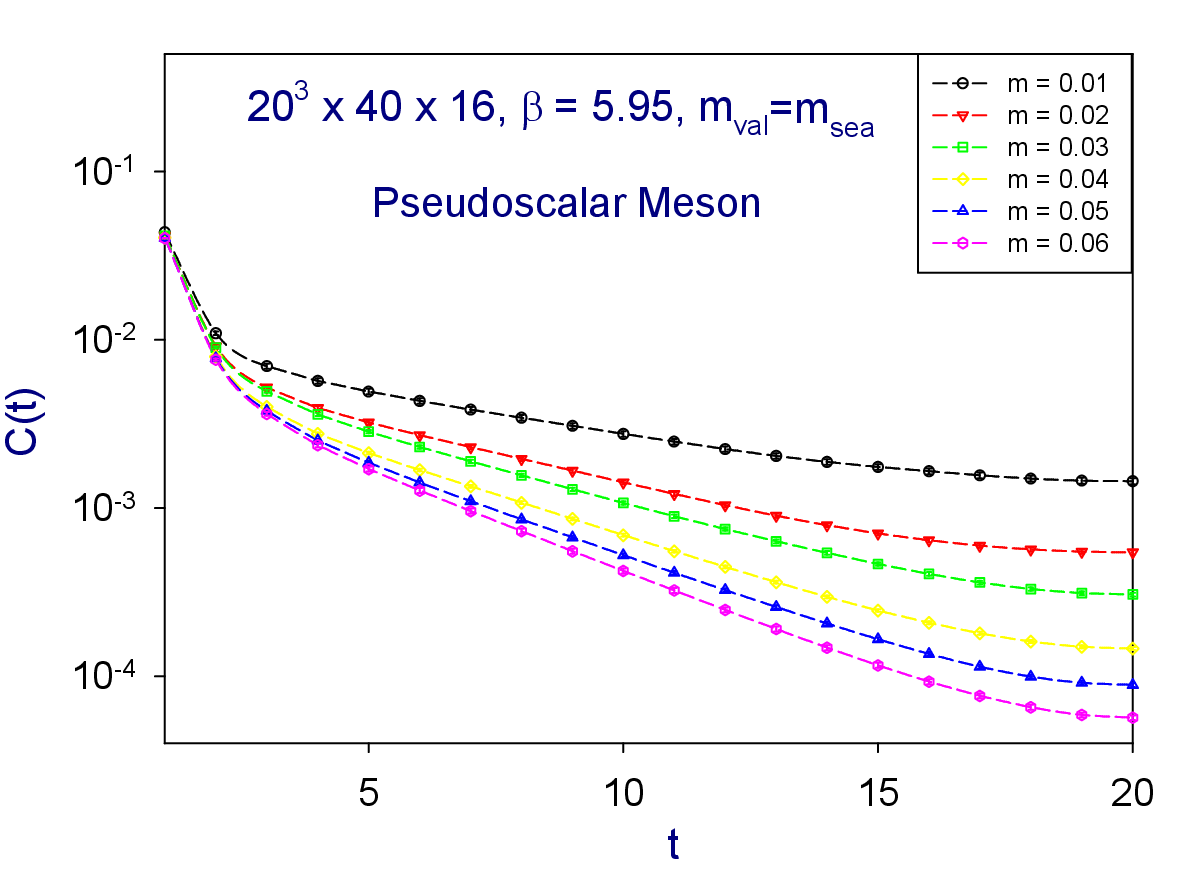}
&
\includegraphics*[width=7.5cm,clip=true]{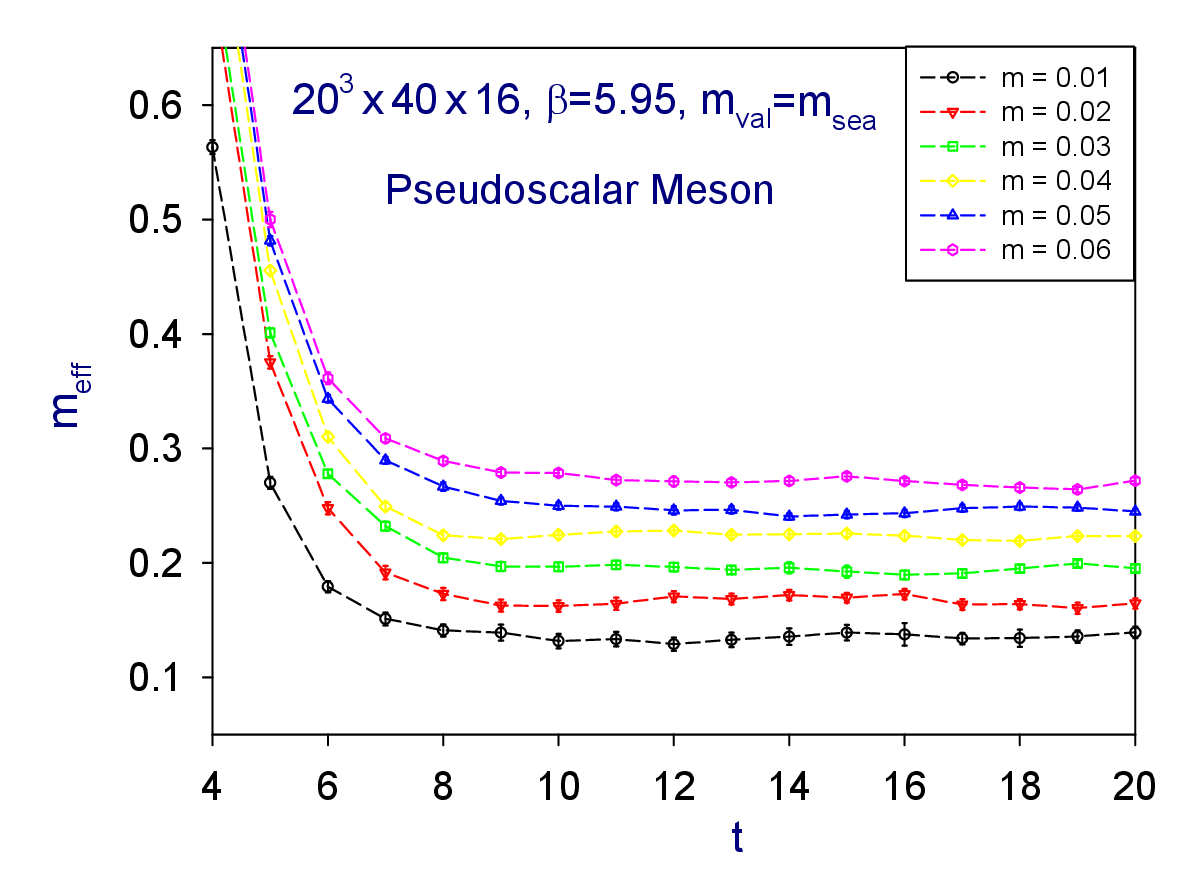}
\\ (a) & (b)
\end{tabular}
\caption{(a) The time-correlation function of the pseudoscalar meson for six sea-quark masses.
                        (b) The effective mass of (a).
         The dashed line connecting the data points of the same sea-quark mass is for guiding the eyes.}
\label{fig:G55a_meff_b595_nf2}
\end{center}
\end{figure*}

\begin{figure}[tb]
\begin{center}
\begin{tabular}{@{}c@{}c@{}}
\includegraphics*[width=7.0cm,clip=true]{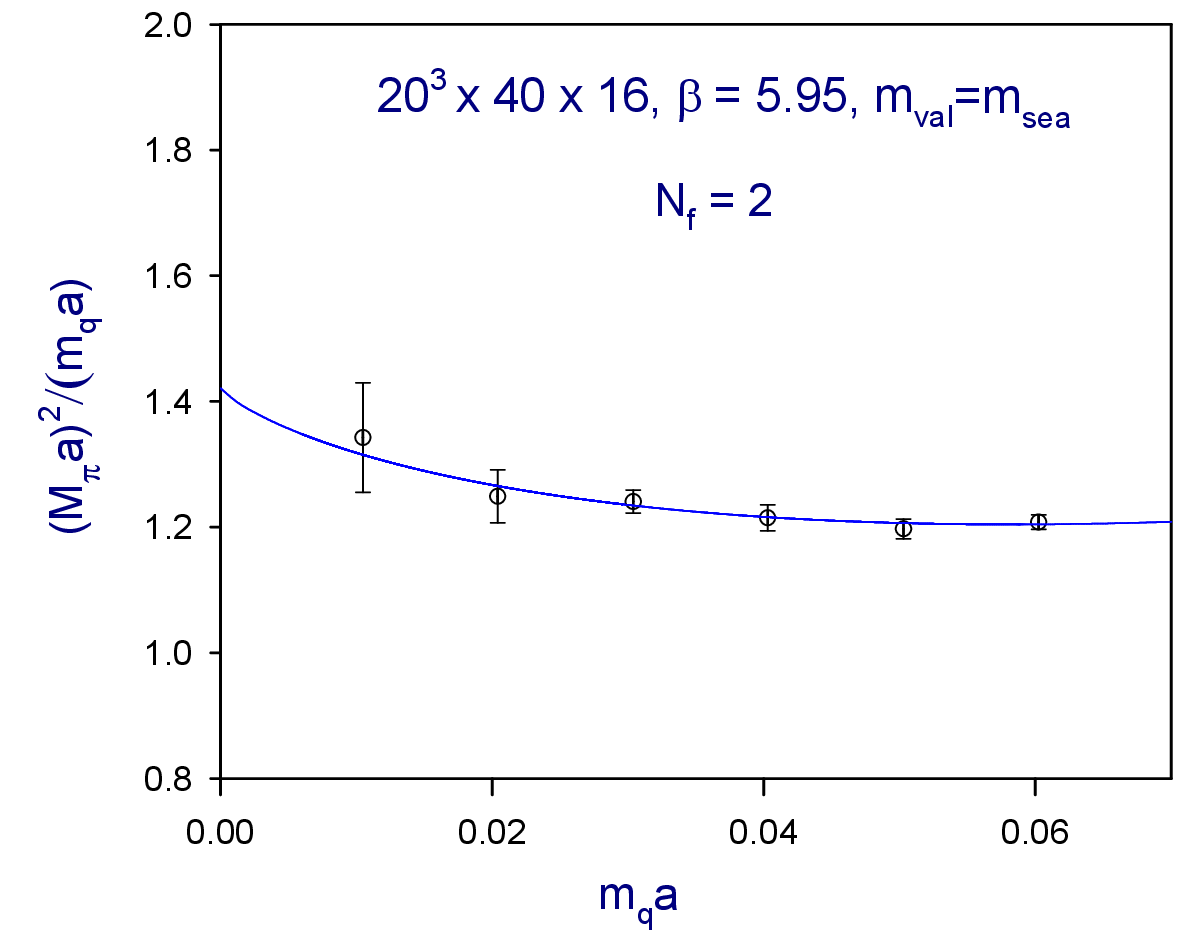}
&
\includegraphics*[width=7.0cm,clip=true]{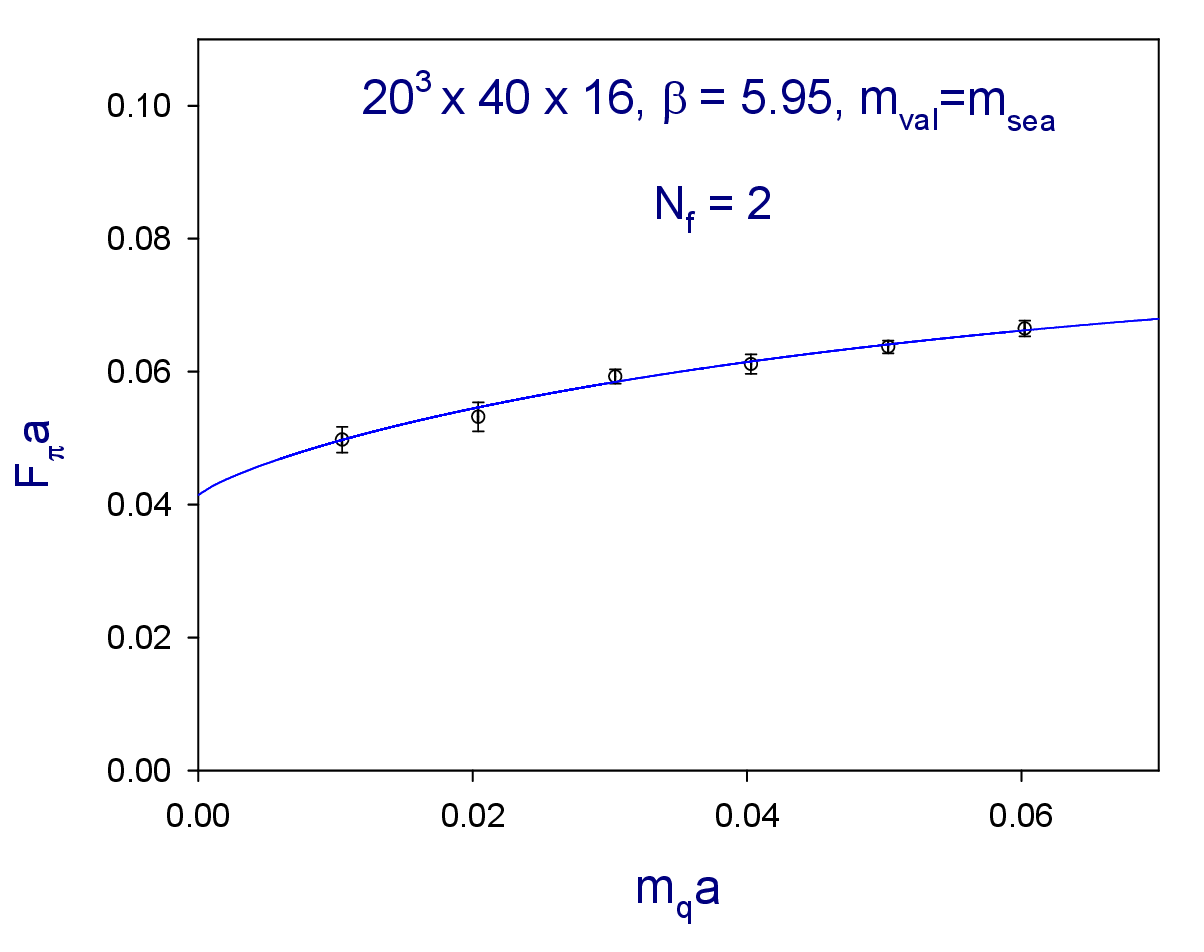}
\\ (a) & (b)
\end{tabular}
\caption{Pseudoscalar meson of 2 flavors QCD with ODWF 
         (a) $ (M_\pi a)^2/(m_q a) $ and (b) $ F_\pi a $.
         The solid lines are the simultaneous fits to the NLO ChPT, for six sea-quark masses.}
\label{fig:mpi2omq_fpi_b595_nf2}
\end{center}
\end{figure}

In figure~\ref{fig:mpi2omq_fpi_b595_nf2},
we plot $ (M_\pi a)^2 /(m_q a) $ and $ F_\pi a $ versus $ m_q a $ respectively.
Here we have made the correction for the finite volume effect 
using the estimate within ChPT calculated up 
to $ {\cal O}(M_\pi^4/(4 \pi F_\pi)^4 ) $ \cite{Colangelo:2005gd}. 
Taking into account of the correlation between $ M_\pi^2/m_q $ and $ F_\pi $ for the same sea-quark mass,  
we fit our data to the formulas of NLO ChPT,
\bea
\label{eq:mpi2omq_NLO_Nf2}
\frac{M_\pi^2}{m_q} &=& \frac{2 \Sigma}{F^2}  \left[ 1 
+ \left(\frac{\Sigma m_q }{16 \pi^2 F^4}\right) \ln\left(\frac{2 \Sigma m_q}{F^2 \Lambda_3^2} \right) \right], \\
\label{eq:fpi_NLO_Nf2}
F_\pi &=& F \left[ 1 -  \left(\frac{\Sigma m_q}{8 \pi^2 F^4 } \right) \ln \left( \frac{2 \Sigma m_q}{ F^2 \Lambda_4^2} \right) \right], 
\eea
where $ \Lambda_3 $ and $ \Lambda_4 $ are related to the low energy constants $ \bar l_3 $ and $ \bar l_4 $ as follows.   
\BAN
\bar l_3 = \ln \left( \frac{\Lambda_3^2}{m_{\pi^{\pm}}^2} \right), \quad
\bar l_4 = \ln \left( \frac{\Lambda_4^2}{m_{\pi^{\pm}}^2} \right), \quad
m_{\pi^{\pm}} = 0.140 \mbox{ GeV}.    
\EAN

Our procedure of data fitting to extract the parameters ($ \Sigma $, $ F $, $ \Lambda_3 $ and $ \Lambda_4 $)
has been outlined in~\cite{Chiu:2011bm}.
For six sea-quark masses,  
our fit gives 
\bea
\begin{array}{l@{\hspace{5mm}}l}
\Sigma a^3 = 0.00122(6)(2), 
&
F a = 0.0414(10)(16), 
\\
\bar l_3 = 3.829(105)(43),
&
\label{eq:l4}
\bar l_4 = 4.755(93)(23),  
\end{array}
\eea   
where the systematic errors are estimated by varying the number of data points from 
6 to 4 ($ m_q a \le 0.04 $ ).

With the fitted parameters, we use the physical ratio  
\BAN
\label{eq:phys_ratio}
\left( \frac{M_\pi}{F_\pi} \right)^{phys} = \frac{0.135 \mbox{ GeV}}{0.093 \mbox{ GeV}} \simeq 1.45 
\EAN 
as the input, and solve the equation $ M_\pi(m_q) / F_\pi(m_q) = 1.45 $
to obtain the physical bare quark mass $ m_q^{phys} a = 0.00254(10)(16)  $. 
From (\ref{eq:fpi_NLO_Nf2}) and the physical pion decay constant $ F_\pi = \mathrm{92.6~MeV} $, 
we determine the inverse lattice spacing at the physical point, 
\BAN    
1/a = \mathrm{2.076(6)(5)~GeV}.  
\EAN
From (\ref{eq:mpi2omq_NLO_Nf2}), we obtain the pion mass at the physical point,  
$M_\pi = 0.134(5)(3) \mbox{ GeV} $, which serves as a consistency check. 

In order to convert the chiral condensate $ \Sigma $ and 
the average $ m_u $ and $ m_d $ to those 
in the $\overline{\mathrm{MS}}$ scheme, 
we calculate the renormalization factor 
$Z_s^{\overline{\mathrm{MS}}}(\mathrm{2~GeV}) $ 
using the non-perturbative renormalization technique
through the RI/MOM scheme \cite{Martinelli:1994ty}, 
which gives
$ Z_s^{\overline{\mathrm{MS}}}(\mathrm{2~GeV}) = 1.244(18)(39) $.
Then the values of $ \Sigma $ and the average of $ m_u $ and $ m_d $ are transcribed to
\bea
\label{eq:sigmaMS}
\Sigma^{\overline{{\mathrm{MS}}}}(\mbox{2 GeV}) = [238(10)(6) \mbox{ MeV}]^3,  \\
\label{eq:mudMS}
m_{ud}^{\overline{{\mathrm{MS}}}}(\mbox{2 GeV}) = 4.07(13)(12) \mbox{ MeV}. 
\eea
Our results of the chiral condensate (\ref{eq:sigmaMS}) and 
the average up and down quark mass (\ref{eq:mudMS})
are in good agreement with our previous ones 
on the $ 16^3 \times 32 $ lattice \cite{Chiu:2011bm}.
Since our calculation is done at a single lattice spacing
the discretization error cannot be quantified reliably, but
we do not expect much larger errors because our lattice
action is free from $O(a)$ discretization effects.

\begin{figure}[!htb]
\begin{center}
\begin{tabular}{@{}cccc@{}}
\includegraphics*[height=5cm,width=4.4cm,clip=true]{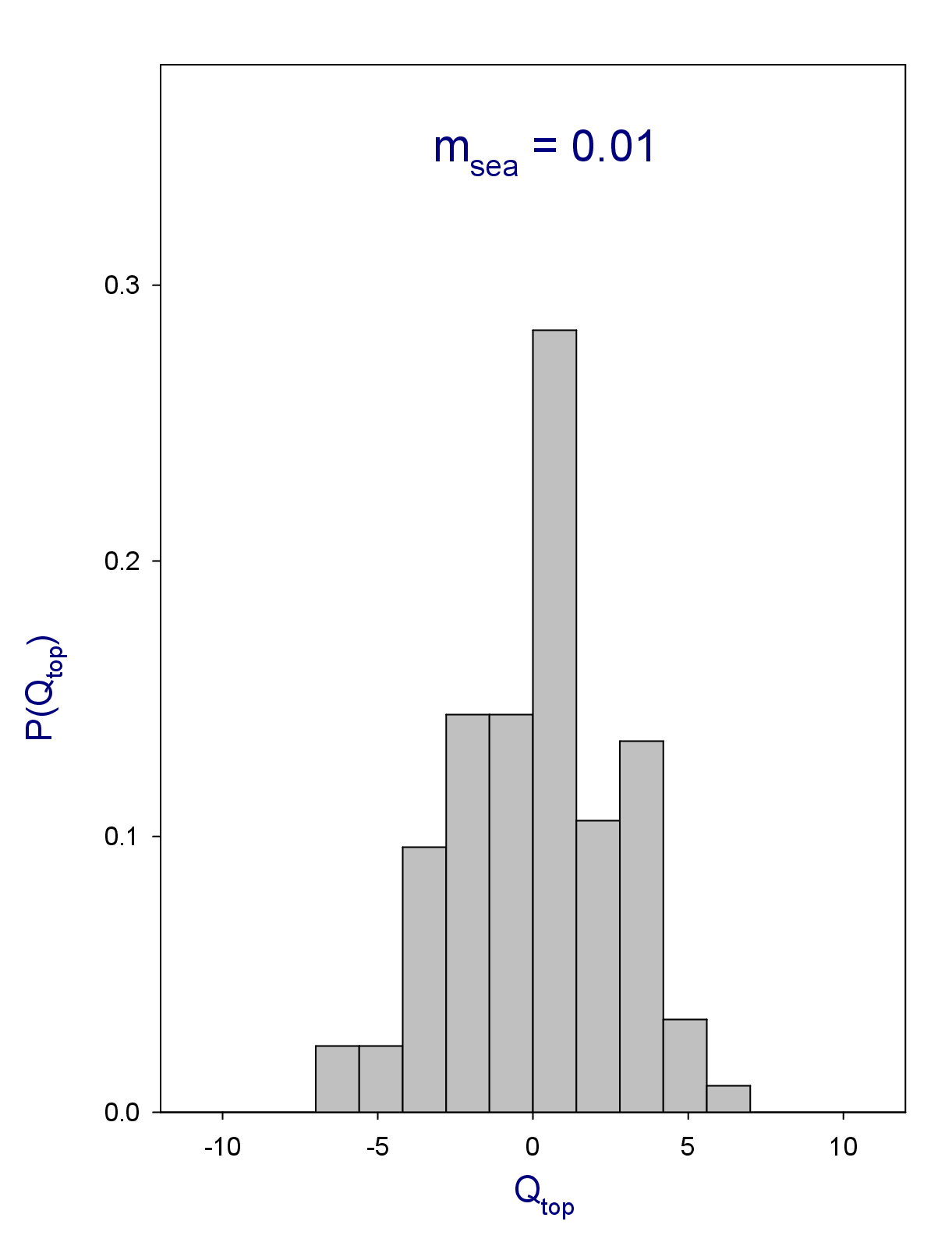}
&
\includegraphics*[height=5cm,width=4.4cm,clip=true]{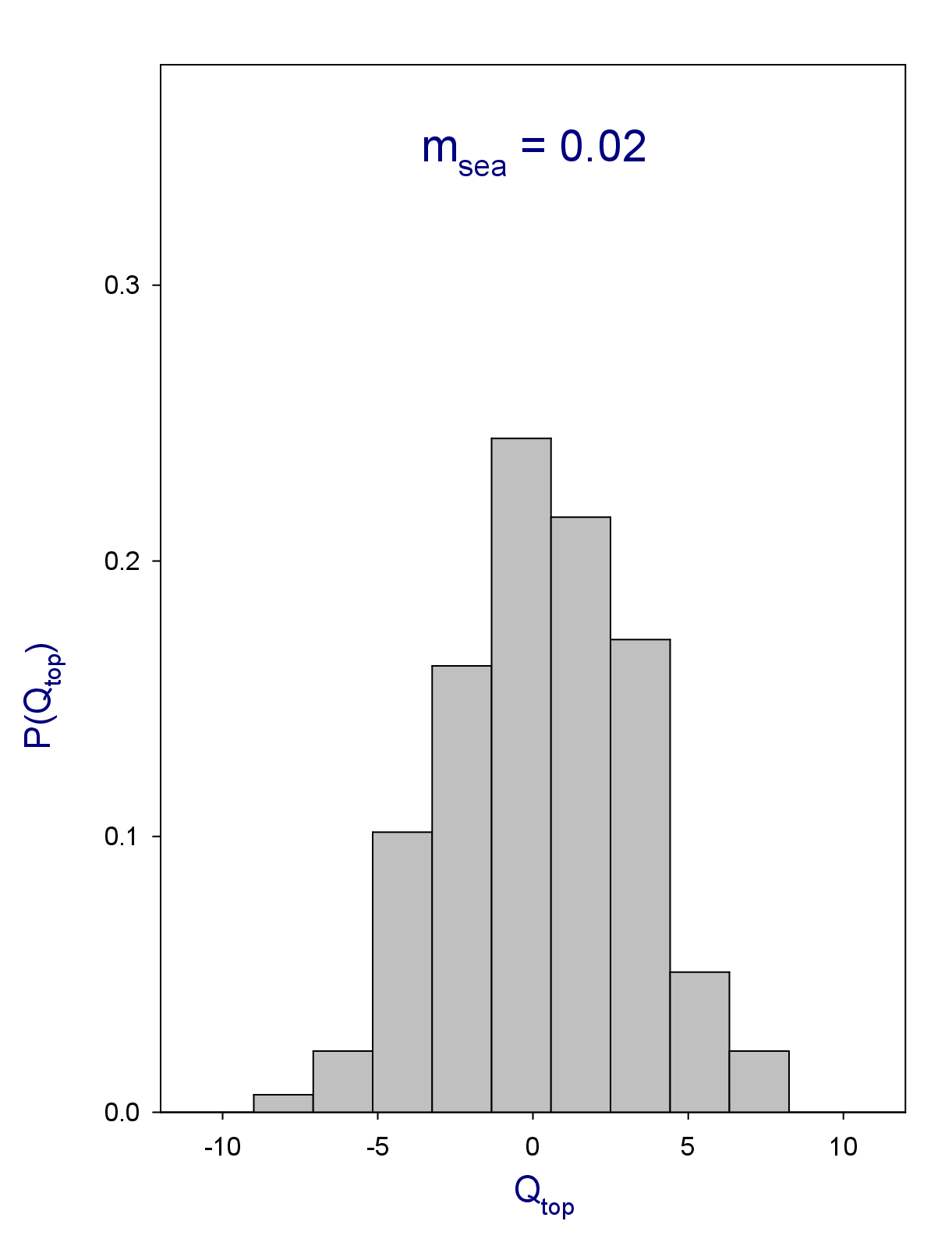}
&
\includegraphics*[height=5cm,width=4.4cm,clip=true]{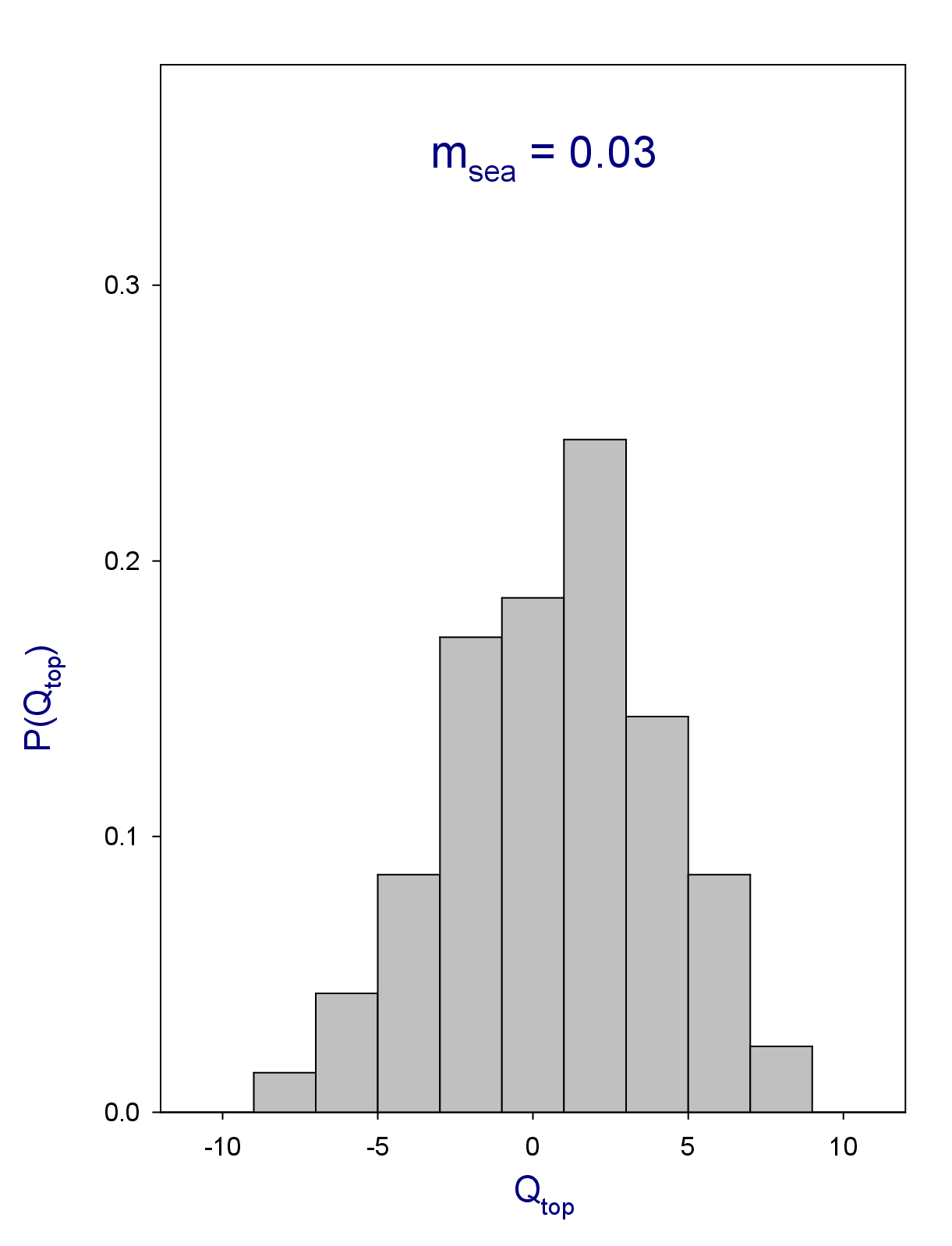}
\\
\includegraphics*[height=5cm,width=4.4cm,clip=true]{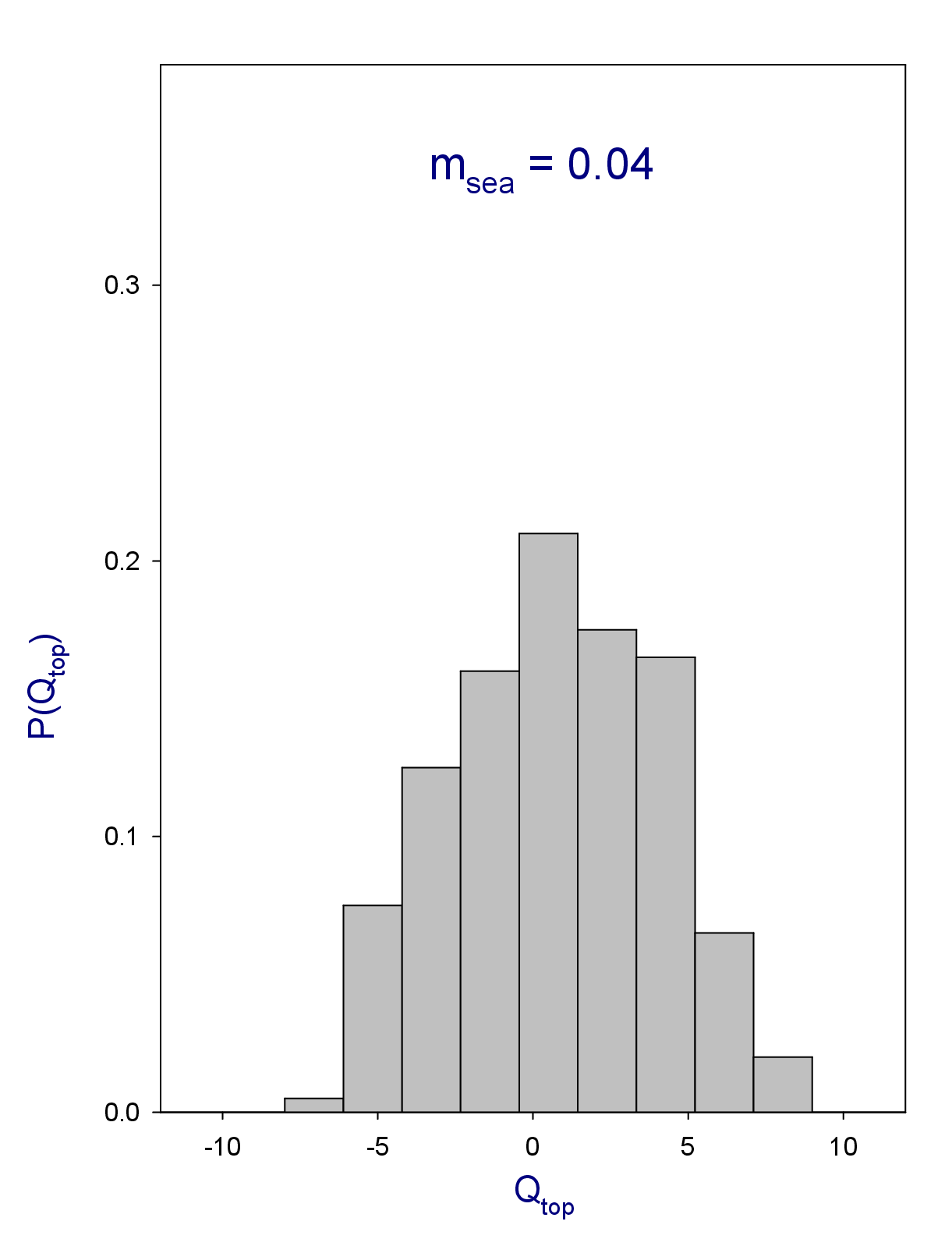}
&
\includegraphics*[height=5cm,width=4.4cm,clip=true]{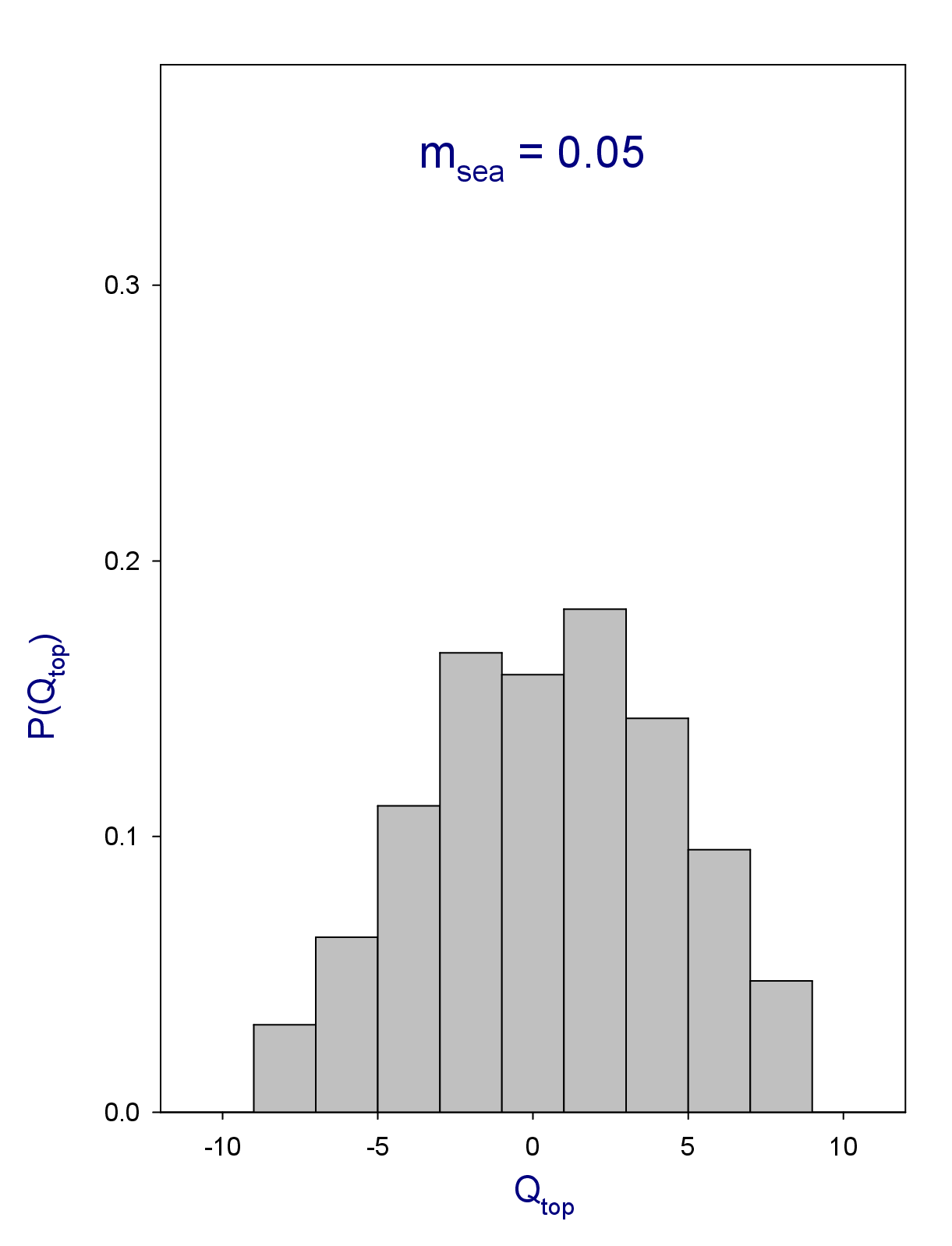}
&
\includegraphics*[height=5cm,width=4.4cm,clip=true]{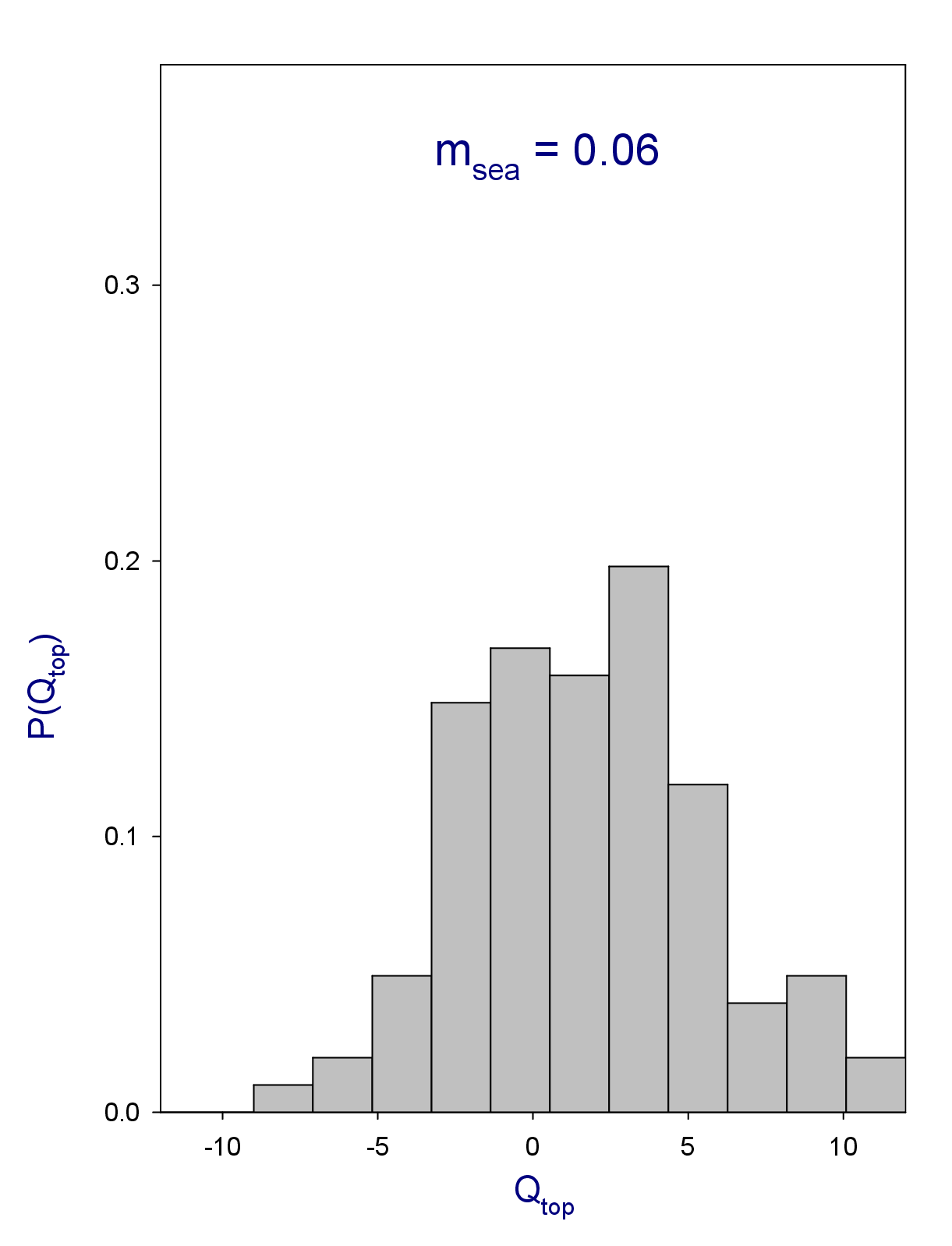}
\\
\end{tabular}
\caption{Histogram of topological charge distribution
for six sea-quark masses (preliminary results with $\sim 200 $ configurations for each ensemble).
}
\label{fig:Q_hist}
\end{center}
\end{figure}

For the projection of low-lying eigenmodes (zero modes plus 180 pairs of lowest-lying eigenmodes)
of the overlap operator for each configuration, we have completed about half of the
total ($ 500 \times 6 = 3000 $) configurations.
In figure~\ref{fig:Q_hist}, we plot the histogram of topological charge distribution
for $ m_q a = 0.01, 0.02, \cdots, 0.06 $ respectively. Evidently, the
probability distribution of $ Q_t $ for each sea-quark mass behaves like
a Gaussian distribution, and it becomes more sharply peaked around $ Q_t = 0 $ as
the sea-quark mass $ m_q $ gets smaller.
We will measure the topological susceptibility and related physical quantities
after the projections of all 3000 configurations are completed.

\section{Conclusion and outlook}

Chiral symmetry plays an important role in particle physics.
It is one of the most salient features of the massless fermion field, 
a consequence of the principles of quantum mechanics and special relativity.
Thus it is vital to preserve the chiral symmetry at finite lattice spacing. 
Theoretically, this is realized by the domain-wall fermions 
with infinite extent ($ N_s \to \infty $) in the fifth dimension, 
or the overlap fermion in the four dimension. 
However, in practice, it is a rather challenging problem to perform 
dynamical simulations of lattice QCD with overlap or domain-wall fermions
such that the chiral symmetry is preserved at a high precision 
and all topological sectors are sampled ergodically. 
Naively, one might expect that a high precision of chiral symmetry
can be attained with a sufficiently large $ N_s $.     
However, the relevant question is how the chiral symmetry violation 
(e.g., the residual mass) decreases with respect to $ N_s $. 

In this work, it has been shown that lattice QCD with ODWF 
provides a viable approach to perform large-scale simulations of unquenched QCD, 
which not only preserves the chiral symmetry 
to a good precision, but also samples all topological sectors ergodically. 
The upper-bound of the residual mass for lattice QCD with ODWF \cite{Chen:2012jya}  
asserts that for $ N_s $ less than some threshold value ($\sim 16-18$), 
the residual mass decays exponentially with respect to $ N_s $. 
Thus, for lattice QCD with ODWF, chiral symmetry can be preserved 
to a good precision ($ m_{res} a \sim 10^{-5} $) with a modest $ N_s \sim 16 $.    
On the other hand, for lattice QCD with the conventional DWF, 
the residual mass behaves like $ N_s^{-\alpha}\; (\alpha \sim 1-2)$ for any $ N_s $ \cite{Antonio:2008zz}.
Thus it is difficult for the conventional DWF to attain 
$ m_{res} a \sim 10^{-5} $, even for $ N_s \sim 32 $. 
Another approach to avoid large chiral symmetry violations is to use 
smeared links rather than thin links, since the number of low-lying eigenmodes 
of $ H = c H_w ( 1 + d \gamma_5 H_w)^{-1} $ would be largely reduced for smeared links. 
However, it is unclear to what extent the short-distance physics would be affected 
even if the smeared links are only used in the lattice fermion operator.

Finally, we briefly outline the new project of the TWQCD Collaboration.    
Now, after simulating two-flavors QCD on various lattices 
($ 16^3 \times 32 $, $ 20^3 \times 40 $, $ 24^3 \times 48 $), 
we are ready to perform dynamical simulations of $(2+1)$-flavors 
and $(2+1+1)$-flavors QCD on the $ 32^3 \times 64 $ lattice, with pion mass close 
to the physical value, and residual mass $ m_{res} a \sim 10^{-5} $.  
We outline our strategy as follows. 
In order to compute the fermion force (by conjugate gradient) for lattice QCD with DWF 
on the $ 32^3 \times 64 \times 16 $ lattice, at least 11 GB RAM are necessary and this exceeds 
the maximum memory (6 GB) currently available in a single GPU (Nvidia C2070/K20x/GTX-TITAN). 
In other words, we must use multiGPUs to meet the memory requirement, 
as well as to speed up the computation. Recently, we have developed efficient 
CUDA codes for the computation of entire HMC trajectories with multiGPUs. 
In table \ref{tab:benchmark}, we summarize our benchmark for various Nvidia GPUs \cite{Chiu:2013gp}.
This suggests that it is feasible to perform dynamical simulations of $(2+1)$-flavors and 
$(2+1+1)$-flavors QCD on the $ 32^3 \times 64 \times 16 $ lattice, with a GPU cluster of 
Nvidia K20/C2075/GTX680/GTX-TITAN. Details will be presented in a forthcoming paper.

\begin{table}[th]
\caption{Benchmark of Nvidia GPUs, using HMC simulation of lattice QCD with ODWF on the $ 32^3 \times 48 \times 16 $ lattice. 
All numbers are in units of Gflops/sec.}
\begin{center}
\begin{tabular}{lllll}
\br
          & 2*C2070 & 2*K20c & 4*GTX680 & 2*GTX-TITAN \\
\mr
          & 340   &  535  &  945  &  774  \\
\br
\end{tabular}
\end{center}
\label{tab:benchmark}
\end{table}

\section*{Acknowledgments}
  I thank the members of the TWQCD Collaboration for many discussions, 
  in particular, Yu-Chih Chen, Wen-Ping Chen, Han-Yi Chou, Tung-Han Hsieh, and Yao-Yuan Mao.  
  Also, I would like to thank the Organizers of CCP2012 
  for invitation and kind hospitality. 
  This work is supported in part by the National Science Council
  (Nos.~NSC99-2112-M-002-012-MY3) and NTU-CQSE (No.~102R891404).
  I also thank NCHC and NTU-CC for providing facilities to perform 
  part of the calculations.

\end{document}